\documentclass[article,aps]{revtex4}

\usepackage{graphicx}
\usepackage{bm}
\usepackage{color}

\newcommand{\be}{\begin{equation}}
\newcommand{\ee}{\end{equation}}
\newcommand{\bea}{\vspace{0.25cm}\begin{eqnarray}}
\newcommand{\eea}{\end{eqnarray}}

\def\PLA{{Phys. Lett.}  A }
\def\PRL{{Phys. Rev. Lett.} }
\def\PRA{{Phys. Rev.} A }

\def\PRD{{Phys. Rev.} D }

\def\PLB{{Phys. Lett.}  B }

\begin{document}
\vskip 0.2cm
\title{  Review of recent experimental progresses in
 Foundations of Quantum Mechanics   and
Quantum Information  obtained in Parametric Down Conversion
Experiments at IENGF. }

\author{  M. Genovese} \email{genovese@ien.it} \author{ G.Brida }\author{ E. Cagliero }\author{ M.
Gramegna}  \affiliation{ Istituto Elettrotecnico Nazionale Galileo
Ferraris, Strada delle Cacce 91, 10135 Torino, Italy }
\author{ E. Predazzi}\affiliation { Dip.
Fisica Teorica Univ. Torino and INFN, via P. Giuria 1, 10125
Torino, Italy }

\begin{abstract}

 We review some recent experimental progresses concerning
Foundations of Quantum Mechanics and Quantum Information obtained
in  Quantum Optics Laboratory "Carlo Novero" at IENGF.

More in details, after a short presentation of our polarization
entangled photons source (based on precise superposition of two
Type I PDC emission) and of the results obtained with it, we
describe an innovative double slit experiment where two degenerate
photons produced by PDC are sent each to a specific slit. Beyond
representing an interesting example of relation between visibility
of interference and "welcher weg" knowledge, this configuration
has been suggested for testing de Broglie-Bohm theory against
Standard Quantum Mechanics. Our results perfectly fit SQM results,
but disagree with dBB predictions.

Then, we discuss a recent experiment addressed to clarify the
issue of which wave-particle observables are really to be
considered when discussing wave particle duality. This experiments
realises the Agarwal et al. theoretical proposal , overcoming
limitations of a former experiment.

Finally, we hint to the realization of a high-intensity
high-spectral-selected PDC source to be used for quantum
information studies.

\end{abstract}

\maketitle

\section{Introduction}

Quantum Mechanics (QM) is one of the pillars of modern physics,
verified by a huge amount of extremely precise experimental data.
Nevertheless, one century after its birth,  it still presents many
unclarified issues at its very foundations as the transition from
a probabilistic microscopic world to a deterministic macroscopic
one, quantum non-locality, the correct interpretation of duality
principle, etc. Of course a clear understanding of all these
points is of the greatest relevance.

Most of foundational problems of QM and its unintuitive properties
are related to {\it entanglement}, which, according to E.
Scr\"odinger, is  "the characteristic trait of quantum mechanics".
For example one can recall the EPR paradox and all the discussion
about Bell inequalities \cite{Aul}.

Beyond its huge conceptual interest, in the last years
entanglement has been recognized as a main resource for quantum
communication and quantum computation \cite{NC}. In quantum
communication the use of entangled photons is, without any doubts,
the main resource for future developments \cite{NC}, being the
basis of various protocols as quantum teleportation \cite{tele},
quantum dense coding \cite{dc} and entanglement swapping \cite{sw}
or quantum key distribution \cite{lontano}. Furthermore, it has
some interest for quantum metrology as well \cite{metr}.

Thus the realization of efficient sources of entangled states is
of the utmost relevance.

In the following we describe our high intensity source of
polarisation entangled states of photons and some application of
it. Then we describe a first test of dBB theory against Standard
Quantum Mechanics obtained in a double slit experiment and discuss
an experiment addressed to reach a conclusive answer to the
interpretation of complementarity principle. Finally, we hint to
the realization of a PDC source of correlated photons with high
spectral selection addressed to an experiment of photon-photon
interaction in Kerr cell.

\section{Our polarisation entangled biphoton source}

Various schemes have been proposed for generating entangled states
of photons.

After the very low efficiency first schemes based on the use of
polarisation entangled states of photons generated by cascade
decay of atoms \cite{asp}, in the 90's a big progress has been
obtained by using parametric down conversion (PDC) processes \cite
{Mandel}. The first experiments \cite{type1,ou} used beam
splitters for generating entanglement, reducing in this way the
total efficiency. More recently,  high intensity sources of
polarisation entangled states of two photons have been obtained by
using Type II PDC \cite{type2} or by superimposing two type I PDC
emissions, in this case either of two thin adjacent crystals
\cite{whi} or by inserting them in an interferometer \cite{shih}.

In our set-up two \cite{nos} type I emissions are also
superimposed, but by using an optical element between the two
crystals. The advantages of this scheme are that, in principle, a
perfect superposition can be obtained and that non-maximally
entangled states of different degrees of entanglement can be
easily realized.

More in details (see Fig.1),  two crystals of $LiIO_3$ (10x10x10
mm) \footnote{Which we have measured by an innovative technique to
have $d_{31} = 3.5 \pm 0.4$ \cite{BGN}.} ( see Fig.\ref{fig:bell})
are placed along the pump laser propagation (an argon laser
emitting at 351 nm), 250 mm apart,  a distance smaller than the
coherence length of the pumping laser. This guarantees
indistinguishability in the creation of a couple of photons in the
first or in the second crystal. An optical condenser, drilled for
leaving pass the pump beam, focalizes PDC emission from the first
crystal into the second one, whose optical axis is rotated of
$90^o$ respect to the first.

A quartz plate and a half-wavelength plate inserted on the path of
pump beam between the two crystals compensate the birefringence on
the first crystal and rotate the polarisation of the beam.

Due to the coherent superposition of the two PDC emissions of
different polarisation the output of this set-up is the state: \be
\vert \Psi \rangle ={\vert H \rangle \vert H \rangle + f \vert V
\rangle \vert V \rangle \over \sqrt{1 + \vert f \vert ^2}}
\label{PsiH} \ee where $H$ and $V$ are horizontal and vertical
polarisation respectively.

A very interesting degree of freedom of this configuration is given by the fact that by tuning the pump intensity between the two crystals one can easily
tune (keeping the maximum available power for the first PDC) the value of $f$,
which determines how far from a maximally entangled state ($f=1$) the produced state is. This is a fundamental property, which allows to select the most appropriate state for the experiment.

This source is very bright since we observe about 10 kHz
coincidence rate at 200 mW pump power (result next to the best
obtained with two adjacent thin type I crystals and by far larger
than the ones realised with type II sources \cite{kw}).

Finally, let us notice that by modifying the polarisation of one
branch or the phase of the parameter $f$ all the four Bell states
can be easily generated.

A test of the state produced by this source can be obtained by measuring Bell inequalities,
as discussed in the following paragraph.

\section{Applications to test Local Realism}

In 1964 Bell demonstrated \cite{bell} that one can test, with complete generality, standard quantum mechanics (SQM) against every local hidden variable theory (LHVT) by considering specific inequalities involving correlation measurement on entangled states. More recently these inequalities assumed a relevant role in quantum communication as well, since it was shown that they can be used for checking the presence of an eavesdropper in quantum key distribution protocols using entangled states \cite{NC}.

Many experiments have already been devoted to test Bell
inequalities \cite{Mandel,asp,franson,type1,type2,whi}, leading to
a substantial agreement with quantum mechanics and disfavouring
realistic local hidden variable theories. However, due to the low
total detection efficiency (the so-called "detection loophole") no
experiment has yet been able to exclude definitively realistic
local hidden variable theories, for it is necessary a further
additional hypothesis \cite{santos}, stating that the observed
sample of particles pairs is a faithful subsample of the whole.
This problem is known as  detection or efficiency loophole.
 Incidentally,  it must be
noticed that a recent experiment \cite{Win} based on the use of Be
ions has reached very high efficiencies (around 98 \%), but in
this case the two subsystems (the two ions) are not really
separated systems during the measurement and the test cannot be
considered a real implementation of a detection loophole free test
of Bell inequalities, even if it represents a relevant progress in
this sense. Analogously, the suggestion that a loophole free
experiment could be obtained by using K or B mesons \cite{BK} has
been shown to be wrong \cite{nosK}, since the detection loophole
reappears due to the necessity of selecting specific decay
channels in order to tag the mesons.

  Considering the extreme relevance
of a conclusive elimination of local hidden variable theories, the research
for new experimental configurations able to overcome the detection loophole
is of the greatest interest.

A very important theoretical step in this direction has been
achieved recognising that for non maximally entangled pairs a
total efficiency larger than 0.67 \cite{eb} (in the limit of no
background) is required to obtain an efficiency-loophole free
experiment, whilst for maximally entangled pairs this limit rises
to 0.81 (see Fig. \ref{f}).

Our set-up, allowing the generation of a chosen non-maximally
entangled state, allows a step in this direction. In particular we
have produced a state with $f \simeq 0.4$.

As a first check of our apparatus, we have measured the interference
fringes, varying the setting of one of the polarisers, leaving the other
fixed. We have found a high visibility, $V=0.98 \pm 0.01$, confirming that a good alignment was reached.

As Bell inequality test we have considered the Clauser-Horne sum,

\begin{equation}
CH=N(\theta _{1},\theta _{2})-N(\theta _{1},\theta _{2}^{\prime
})+N(\theta _{1}^{\prime },\theta _{2}) \\ +N(\theta _{1}^{\prime
},\theta _{2}^{\prime })-N(\theta _{1}^{\prime },\infty )-N(\infty
,\theta _{2})  \label{eq:CH}
\end{equation}
which is strictly negative for local realistic theory. In
(\ref{eq:CH}), $N(\theta _{1},\theta _{2})$ is the number of
coincidences between channels 1 and 2 when the two polarisers are
rotated to an angle $\theta _{1}$ and $\theta _{2}$ respectively.
Because of low detection efficiency we have substituted in Eq.
\ref{eq:CH}, as in any experiment performed up to now,  single
counts $N(\theta _{1}^{\prime } )$ and $N(\theta _{2})$ with
coincidence counts $N(\theta _{1}^{\prime },\infty )$ and
$N(\infty ,\theta _{2})$, where $\infty $ denotes the absence of
selection of polarisation for that channel. This is one of the
form in which detection loophole manifests itself.

On the other hand,  quantum mechanics predictions for $CH$ can be
larger than zero: for a maximally
entangled state the largest value is obtained for $\theta _{1}=67^{o}.5$ , $%
\theta _{2}=45^{o}$, $\theta _{1}^{\prime }=22^{o}.5$ , $\theta
_{2}^{\prime }=0^{o}$ and corresponds to a ratio
\begin{equation}
R=[N(\theta _{1},\theta _{2})-N(\theta _{1},\theta _{2}^{\prime
})+N(\theta _{1}^{\prime },\theta _{2})+N(\theta _{1}^{\prime
},\theta _{2}^{\prime })]/[N(\theta _{1}^{\prime },\infty
)+N(\infty ,\theta _{2})]  \label{eq:R}
\end{equation}
equal to 1.207.

For non-maximally entangled states the angles for which CH is
maximal are somehow different and the maximum is reduced to a
smaller value. The angles corresponding to the maximum can be
evaluated maximising Eq. \ref{eq:CH} with

\bea \left. \begin{array}{l}

  N[\theta _{1},\theta _{2}] =  [ \epsilon _1^{||} \epsilon _2^{||} (Sin[\theta _{1}]^{2}\cdot Sin[\theta_{2}]^{2}) + \\
  \epsilon _1^{\perp} \epsilon _2^{\perp}
(Cos[\theta _{1}]^{2} \cdot Cos[\theta _{2}]^{2} )\\
  (\epsilon _1^{\perp} \epsilon _2^{||} Sin[\theta _{1}]^2\cdot Cos[\theta _{2}]^2 + \epsilon _1^{||} \epsilon _2^{\perp}
Cos[\theta _{1}]^2 \cdot Sin [\theta _{2}]^2 )  \\
 + |f|^{2}\ast (\epsilon _1^{\perp} \epsilon _2^{\perp} (Sin[\theta _{1}]^{2}\cdot Sin[\theta_{2}]^{2}) +  \epsilon _1^{||} \epsilon _2^{||}
(Cos[\theta _{1}]^{2} Cos[\theta _{2}]^{2} ) +\\
(\epsilon _1^{||} \epsilon _2^{\perp} Sin[\theta _{1}]^2\cdot Cos[\theta _{2}]^{2} +\\
 \epsilon _1^{\perp} \epsilon _2^{||}
Cos[\theta _{1}]^2 \cdot Sin [\theta _{2}]^2 )   \\
  +  (f+f^{\ast }) ((\epsilon _1^{||} \epsilon _2^{||} + \epsilon _1^{\perp} \epsilon _2^{\perp} - \epsilon _1^{||} \epsilon _2^{\perp} -
\epsilon _1^{\perp} \epsilon _2^{||}) \cdot (Sin[\theta _{1}]\cdot
Sin[\theta _{2}]\cdot Cos[\theta _{1}]\cdot Cos[\theta _{2}]) ]
/(1+|f|^{2}) \,
\end{array}\right. \, .
\label{cc} \eea where (for the case of non-ideal polariser)
$\epsilon _i^{||}$ and $\epsilon _i^{\perp}$ correspond to the
transmission when the polariser (on the branch $i$)  axis is
aligned or normal to the polarisation axis respectively.

The  phase of $f$ must be kept next to zero. Any relative phase
between the two components of the entangled state reflects into a
reduction of Clauser-Horne inequality violation, up to reaching no
violation at all for a phase difference of $\pi /2$.

In order to measure $CH$ we selected  two conjugated directions at
789 and 633 nm geometrically and by using two interference filters
of 4 nm FWHM preceding the avalanche photodiode photo-detectors.

For our produced state, corresponding to $f \simeq 0.4$,  the
largest violation of the inequality is reached for $\theta_1
=72^o.24$ , $\theta_2=45^o$, $\theta_1 ^{\prime}= 17^o.76$ and
$\theta_2 ^{\prime}= 0^o$, to $R=1.16$.

Our experimental result $CH = 513 \pm 25$ coincidences per second,
is more than 20 standard deviations from zero and compatible with
the theoretical value predicted by quantum mechanics. In terms of
the ratio (\ref{eq:R}), our result is $1.081 \pm 0.006$. The
smaller value respect to the theoretical prediction is easily
explained in terms of a residual non-perfect alignment.

For the sake of comparison, one can consider the value obtained
with the angles which optimize Bell inequalities violation for a
maximally entangled state. The result is $CH = 92 \pm 89$, which,
as expected, shows a smaller violation than the value obtained
with the correct angles setting.

Thus, our result represents a further indication favouring SQM
against LHVT. Its main interest is due to the fact that using
tunable non-maximally entangled states is a relevant step toward a
conclusive experiment eliminating the detection loophole.

Furthermore, it allows also to exclude some specific local
realistic models. In particular, we have considered the model of
Casado et al. \cite{Santos2}. These authors have presented a local
realistic model addressed to be compatible with all the available
experiments performed for testing local realism. This model
represents the completion of series of papers where this scheme
has been developed \cite{Santosv}. The main idea is that the
probability distribution for the hidden variable is given by the
Wigner function, which is positive for photons experiments.
Furthermore a model of photodetection, which departs from quantum
theory,  is built in order to reproduce available experimental
results.

This model has the great merit of giving a number of constraints,
which do not follow from the quantum theory and are experimentally
testable.

In particular, there is a minimal light signal level which may be reliably detected: a difference from quantum theory is predicted at low detection rates, namely when the single detection rate $R_S $ is lower than

\begin{equation}
R_S < { \eta F^2 R_c^2 \over 2 L d^2 \lambda \sqrt{ \tau T} }
\label{rate}
\end{equation}
where $\eta$ is the detection quantum efficiency, F is the focal distance of the lens in front of detectors,
$R_c$ is the radius of the active area of the non-linear medium where entangled photons are generated,
$\tau$ is the coherence time of incident photons,
d is the distance between the non-linear medium and the photo-detectors, $\lambda$ the average wavelength of detected photons. L and T are two free parameters which are less well determined by the theory \cite{Santosv}: L can be interpreted as the active depth of the detector, while T is the time needed for the photon to be absorbed and should be approximately less than 10 ns \cite{Santosv}, being, in a first approximation, the length of the wave packet divided for the velocity of light.

Referring to the parameters of Eq. \ref{rate} we have $\eta = 0.51 \pm 0.02$ (a value which we have directly measured by
 using PDC detector calibration \cite{JMO}), F= 0.9 cm, $R_c = 1 $ mm, d= 0.75 m and $\tau = 4.2 \cdot 10^{-13} s$ (due to spectral selection by an interferential filter). L can be estimated of $3 \cdot 10^{-5}$ m.
This leads to $T > 1 s$, extremely higher than the limit of 10 ns
suggested in the model \cite{Santos2}. Thus, we are strictly in
the condition where quantum mechanics predictions are expected to
be violated and, in particular, a strong reduction of visibility
is expected. Nevertheless, our results show  a strong violation of
Clauser-Horne inequality and a high visibility, in  agreement with
standard quantum mechanics, and therefore substantially exclude
\cite{nosW} the model of Ref. \cite{Santos2} \footnote{For a
further, negative, test of this model see also \cite{SnosS}}.

For the sake of completeness, it can be noticed that one of the
authors of the previous model presented a new LHVT model
\cite{s3}, which does not have the same degree of development of
the former one, but in its simplicity allows to reproduce all Bell
inequalities tests performed with polarisation entangled photons.

In Ref. \cite{s3} it is suggested that a test of the model can be
performed by comparing the visibility:

\be V_a= { N(0) - N(\pi /2) \over N(0) + N(\pi /2)} \ee

with

\be V_b = \sqrt{2} { N(\pi /8) - N(3 \pi /8) \over N(\pi /8) + N(3
\pi /8)} \ee

where $N(\theta)$ are the coincidence counts when the two
polarizers are set to two angles differing of  $\theta$.

In fact, in the model of Ref. \cite{s3} \be V_b / V_a > 1+ cos^2
(\pi \eta / 2) \left [ V_b - { sin^2 ( \pi \eta / 2) \over (\pi
\eta / 2)^2} \right] \label{ins} \ee is expected, result that can
be violated in SQM.

For the moment, the use of our data does not allow to exclude this
model since we obtain for the inequality \ref{ins} the value
$1.177 > 1.04$ (a value above unity probably denotes that our
data, not explicitly taken for this purpose, do not yet allow a
sufficient accuracy for this test). A further dedicated experiment
will be realized in a near future.

Finally, we would like to point out that this high intensity
source can find a natural application in quantum communication. In
particular, at the moment we are implementing a scheme \cite{epjd}
for codifying in four dimensional Hilbert spaces (since
codification in higher dimensional Hilbert Spaces may present a
larger security \cite{qutrit}) and another addressed to use Kerr
interaction in rubidium atoms cell for realizing a controlled
unitary gate at single photon level \cite{Kerr}.

\section{An innovative double slit experiment}

Even if Bell inequalities experiments will lead to a conclusive test of local hidden variable theories, non-local hidden variable models (NLHVT) will still be possible.

The most interesting example of NLHVT is the de Broglie Bohm theory (dBB).
dBB \cite{7} is a deterministic theory where the hidden variable (determining the evolution of a specific system)
is the position of the particle, which follows a perfectly defined trajectory in its motion.
The evolution of the system is given by classical equations of motion, but an additional potential  must be included.
 This "quantum" potential is related to the wave function of the system and thus it is non-local.
  The inclusion of this term, together with an initial distribution of particle positions given by the quantum
  probability density,  successfully allows the reproduction of {\it almost} all the predictions of quantum mechanics.
  Nevertheless, a possible discrepancy between SQM and dBB in specific cases has been recently suggested by  Ref.s \cite{4,5,6}

In particular Ref.s \cite{4,5,6} suggest that differences can appear in a double slit experiment where two
identical particles cross each a specific slit at the same time.

Such a configuration can be easily realized with our set-up
substituting the second crystal with a double slit. In particular
we have used two slits separated by  100  $\mu$m  of a width of 10
$\mu$m. They lay in a plane orthogonal to the incident laser beam
and are orthogonal to the table plane.

Different PDC photon pairs crossing the double slit are
statistically distributed with a dispersion of the order of $10
\mu m$ determined by geometrical acceptance, but photons in a
single pair have strong spatial correlation. By using the
formalism presented in Ref. \cite{joe}, we can calculate that the
dispersion of the positions of photons of a single pair at the
double slit is geometrically around $0.25 \mu m$ (see the Fig.
\ref{slit}). Furthermore, in any case the two photons arrive at
the double slit at the same time, in the sense that both cross the
slit at the same instant largely within their coherence time
($\approx 400 fs$). In this sense the theoretical proposal is well
realized and a partial penetration of at most  few tens of
micrometers of trajectories in the same semiplane is expected.

 Two single photon detectors are placed at 1.21 m and at 1.5 m
from the slits after an interference filter at 702 nm, whose full
width at half height is 4 nm, and a lens of 6 mm diameter and 25.4
mm focal length.

In Fig. \ref{fig:3} we report the measured coincidence pattern.
The data are obtained by averaging  7 points of 30' acquisition
each. One detector is placed at $-5.5$ cm from the symmetry axis,
whilst the second is moved sweeping the whole diffraction peak.
The data are in agreement with the pattern predicted by SQM, \bea
& C(\theta_1,\theta_2) = &  g(\theta _1, \theta _i^A )^2 g(\theta
_2, \theta _i^B)^2 + g(\theta _2, \theta _i^A )^2 g(\theta _1,
\theta _i^B)^2 + \cr & & 2 g(\theta _1, \theta _i^A ) g(\theta _2,
\theta _i^B) g(\theta _2, \theta _i^A ) g(\theta _1, \theta _i^B)
cos [ k s (sin \theta _1 - sin \theta _2)] \eea where \be g(
\theta , \theta _i^l)  = { sin ( k w /2 ( sin (\theta) - sin
(\theta _i^l)) \over k w /2 ( sin (\theta) - sin (\theta _i^l))}
\ee
takes into account diffraction. k is the wave vector, s the
slits separation, w the slit width, $\theta _{1,2}$ is the
diffraction angle of the photon observed by detector 1 or 2,
$\theta_i^l$ the incidence angle of the photon on the slit $l$ (A
or B).

A clear coincidence signal is observed also when the two detectors
are placed in the same semiplane respect to the double slit
symmetry axis. In particular, when the centre of the lens of the
first detector is placed -1.7 cm
 after  the median symmetry axis  of the two slits (the minus means to the left of the symmetry axis looking towards
 the crystal) and the second detector is kept at -5.5 cm, with 35 acquisitions of 30' each we obtained 78 $\pm$ 10
 coincidences per 30 minutes after background subtraction, ruling out a null result at nearly eight standard deviations.
Thus, if  the former theoretical prediction will be confirmed,
this experiment poses a strong constraint on the validity of de
Broglie-Bohm theory, which is the most successful example of a
non-local hidden variable theory, representing a very relevant
progress on the line of a final clarification of foundations of
quantum mechanics \cite{news}.

Even if the former theoretical prediction is still somehow subject
to discussion \cite{disc}, we think that our results
\cite{nosdBB}, in agreement with SQM predictions but at variance
with dBB ones, represent a relevant contribution to the debate
about the foundations of quantum mechanics urging a final
clarification about validity of this theoretical proposal.

A further interesting property of this scheme is that it allows a new clear test of the connection between which path knowledge and absence of interference.
Since idler and signal photons have no precise phase relation and each photon crosses a well defined slit, no interference appears at single photon detection level. When the coincidence pattern is considered, path undistinguishability is established since the photodectector 1 (2) can be reached either by the photon which crossed slit A or by the one that went through slit B and vice versa. Thus, even if no second order interference is expected, a fourth order interference  modulates the observed diffraction coincidence pattern.

In Fig. \ref{fig:4} we report the observed coincidence pattern (with 10 acquisitions of one hour for each point)
obtained when the first detector scans the diffraction pattern, while the second is positioned at $-1$ cm
from the symmetry axis. The iris in front of the first detector is of 2 mm. Even if the data have large uncertainties
there is a good indication of the fourth-order interference: the interference pattern predicted by SQM fits the data
with a reduced $\chi ^ 2$  of $0.9$. By comparison, a linear fit (absence of interference) gives $\chi ^ 2 =12.6$
(with 5 degrees of freedom) and is therefore rejected with a $5 \%$ confidence level. On the other hand we have
checked that, as expected, the single channel signal does not show any variation in the same region: the measured
ratio between the mobile and the fixed detector is essentially constant (within uncertainties) in this region.

\section{Wave particle duality experiment}

Wave particle duality is one of the fundamental  aspects of
Quantum Mechanics. Nevertheless, the original Bohr statement about
the term complementarity (in  particular complementarity between
wave and particle behaviours) "to denote the relation of mutual
exclusion characteristic of the quantum theory with regard to an
application of the various classical concepts and ideas"
\cite{Bohr} has been recently subject of a wide debate and a
paradigm where this "mutual exclusion" must be interpreted in a
weaker sense is emerging.

Several experiments with single photons and atoms in
interferometers have shown how a gradual transition takes place
between the two aspects, wave (interference) and particle (which
path knowledge) \cite{Aul,int}. The knowledge of "welcher weg "
(which path) is therefore alternative to coherence (and thus to
the possibility of having interference) with a smooth transition
between a perfect "welcher weg" knowledge and a $100 \%$
interference visibility. However, the extension of complementarity
to classical concepts of waves and particle in every situation
(including, also, for example tunnel effect or birefringence) is
not contained in the mathematical formalism of Quantum Mechanics
(and not tested by interference experiments) and can be questioned
\cite{PL}. Furthermore, in interference experiments \cite{int,Aul}
the use of beam splitters (or similar devices) can be somehow
modellized with classical particles transmission and reflection
\cite{P1}.

In this sense a large interest arouses an experiment \cite{jap}
based on the theoretical proposal of Ref. \cite{P0} where the
coincidences between photodetectors after a tunnel effect in a
double prism of single photons produced in Parametric Down
Conversion are studied.

In more detail, a single photon arriving on two prisms separated
by a small distance (less than the photon wave length) can either
be totally reflected or tunnels through the gap. In the first case
it will be sent to a first detector, in the second case to another
one; coincidences (anticoincidences) between these two detectors
are then measured. The "sharp" particle, anticoincidence, and
wave, tunnel (rather than interference), properties are
simultaneously realized. The observation of coincidences is
incompatible with quantum optics, but could be explained \cite{P0}
in terms of stochastic optics \cite{santosM}.

The result of this experiment was the observation of
anticoincidences between detectors showing that single photons
both had performed a tunnel (wave behaviour) and had been detected
in only one of the detectors ("whelcher weg" knowledge), leading,
therefore, to an agreement "with quantum optics, namely, light
showed both classical wave-like and particle-like pictures
simultaneously" "in contrast with conventional interpretation of
the duality principle"  \cite{jap}.

In Ref. \cite{P1} this result was interpreted as an indication in
favour of de Broglie-Bohm theory.

Nevertheless, the results of this experiment were questioned
\cite{Un,PL} as a case of "an insufficient statistical precision".
More in detail, in Ref. \cite{int} it was introduced the parameter
$\alpha = {N_c N \over N_1 N_2}$ (where $N_c$ denotes coincidence
counts, $N$ the number of gates where photons are counted and
$N_1$ and $N_2$ single detector counts), which should be $\ge 1$
for a classical source and
 $<< 1$ for PDC quantum states \cite{int} (strictly zero
in absence of background) applying it to a test of interference -
"welcher weg" knowledge experiment. In Ref. \cite{Un} this
parameter was estimated to be $\alpha \simeq 1.5 \pm 0.6$ for the
data of Ref. \cite{jap}. Given that, according to Ref.
\cite{int,Un,PL}, this parameter $\alpha$ is the best
discriminator between classical and quantum states, the
experimental precision of Ref. \cite{jap} was therefore largely
insufficient to discriminate between classical and quantum light.

Considered the large relevance of these studies for the very
foundations of Quantum Mechanics, we have  realized \cite{noswp} a
new version of this experiment, where the wave behaviour is
related to birefringence as suggested in Ref. \cite{PL},
overcoming the previous limitations.

Our scheme consists (see Fig. \ref{fig:1}) of a heralded single
photon source based on type I parametric fluorescence generated by
an UV pump laser (at 351 nm) into a non-linear crystal. From PDC
properties, the observation of a photon (at a 633 nm), after
spatial and spectral selection, in a first detector (D3) implies
the presence of a second photon on the conjugated direction (at
789 nm) and can therefore be used to open a coincidence window (by
starting a ramp of two TACs) where the second photon is expected
to be detected . Before detection this second photon crosses a
birefringent crystal where its path is split according to its
polarisation: birefringence (and in particular the fact that
refractive indices are both larger than unity) is a typical
phenomenon explained only in terms of wave like propagation.
Finally, two single photon detectors (D1 and D2) are placed on the
two possible paths (for ordinary and extraordinary polarisation)
and their outputs are respectively routed (as stop) to the
previous TACs. The measurement of coincidences between these two
last detectors (by an AND circuit) in the window opened after a
count in the first detector (D1) allows \cite{PL}, in complete
analogy to Mizobuchi and Ohtak\'e experiment \cite{jap},
observation of corpuscolar properties of the photon (specific
path) together with wave ones (birefringence). On the other hand,
the use of a high intensity source and of a simple scheme allow to
overcome the low statistic limitations of the previous experiment.

In this configuration the logical AND between the valid starts of
the two TACs (where the start is the number of counts measured by
D3) represents therefore the number of gates (N). $N_c$ and $N$,
together with the number of counts in the 7ns temporal window
measured by D1 ($N_1$) and D2 ($N_2$), allow the evaluation of
$\alpha = {N_c N \over N_1 N_2}$.

The results of our experiment for this parameter  in function of
the average single counts on trigger channel D3 (corresponding to
different attenuations of the pump laser beam), are shown in Fig.
\ref{fig::2}. The data are obtained with 500 acquisitions of 1 s
per point, except the one at 20000 counts/s obtained with 5000
acquisitions of 1 s. As expected, due to small background, data
are compatible with zero at low single counts values. At larger
values accidental random coincidences are not anymore negligible
and the measured $\alpha $ increases, remaining however largely
under unity in  the whole investigated region.
 The weighted average of the first three points $\alpha= 0.022 \pm
 0.019$ is (within almost one standard deviation) compatible with
 zero, as expected from QM, differing from unity of more than 51 standard deviations.

In order to compare PDC single photon results with a classical
source, we have repeated the measurement by using an attenuated
He-Ne laser, emitting at 633 nm, and a thermal source (tungsten
lamp). In this case the gate (the start of TACs) is given by a
pulse generator with a trigger frequency rate of 65 kHz (see
Fig.\ref{fig::3}); the rest of the apparatus is the same as
before. In Fig.\ref{fig::4} (similarly to Fig.\ref{fig::2}) we
report the results (with 500 acquisitions of 1 s per point) for
the laser as function of the average single counts of one of the
detectors (i.e. of laser power, attenuated by inserting neutral
filters). Our experimental average datum $\alpha = 0.9980 \pm
0.0022$ is in perfect agreement (within one standard deviation)
with the result expected  for laser coherent light, i.e.
\cite{int} $\alpha =1$. Similarly, for the thermal source we
obtain (see Fig.\ref{fig::5}) the average value $\alpha = 1.0010
\pm 0.0028$ in perfect agreement with the expected value $\alpha =
1$.

Thus, in conclusion, we have realized a new version of the
experiment suggested in Ref. \cite{P0,PL}, which overcomes
limitations \cite{Un} of a previous similar experiment  \cite{jap}
leading to a conclusive answer to the questions raised in the
original theoretical paper.

\section{Source of PDC photons with a high spectral selection }

In the last years the need of high intensity PDC sources has
assumed a large relevance for several applications.

Among them, interaction of entangled photons with atomic levels,
which requires very strong spectral selection (a band width around
$10^{-5} nm$) and therefore a large initial power for having a
final sufficient intense signal, is very important for many
applications as quantum memories \cite{atom}, remote quantum
clocks synchronisation \cite{or}, realisation of quantum logical
gates, etc. In our laboratory a work is in progress (in
collaboration with Camerino University, LENS and INOA) for
realising a controlled-not gate by interaction of two photons by
mean of Kerr effect in Bose-Einstein condensate \cite{Kerr}.

For this purpose we have realized a PDC pulsed source with a
spectral selection up to 0.01 nm obtained with two monochromators.

The source is based on a 5 ns (10 Hz repetition rate) triplicate
Neodimium-YAG laser (355 nm) emitting pulses with a power up to
200 mJ that are used for pumping a 5x5x5 mm BBO crystal where PDC
light is generated (see Fig. \ref{rub}). After the crystal and a
first spatial selection two conjugated directions, corresponding
to a wavelength of 780 nm and 651 nm, are addressed to the
monochromators.

In order to clearly identify the direction corresponding to the
wavelength 780.251 nm of the rubidium 87 transition 5 $2^S_{1/2}$
F=2 into 5 $2^P_{3/2}$ F'=3 we have locked a diode laser to the
transition line and then used it for tracing the optical paths of
this wavelength and of the conjugated stimulated emission in the
PDC. The locking is achieved by injecting the laser diode beam
into a Rb gas sample and varying its frequency until a dip in the
transmission profile of the Rb sample is seen on a photodiode
output; the frequency modulation is achieved by means of a
modulation in the length of the extended cavity of the laser
diode. The alimentation current of the laser diode is then finely
varied by a feedback electronic module in order to lock the
frequency to the transmission dip corresponding to the chosen Rb
line.

The output of monochromators is then addressed to two avalanche
photo-diodes, whose output is routed to a two channel counter, in
order to have the number of events on single channel, and to a
Time to Amplitude Converter circuit, followed by a single channel
analyser, for selecting and counting coincidence events.

Even if the low repetition rate of the laser (and the dead time of
detectors) limits the number of coincidences to 10 per second, due
to the high intensity of the laser the number of pairs produced
per pulse is high and one has a large number of pairs even after a
strong spectral selection. The set-up is therefore suited for
looking for Electromagnetic-Induced-Transparency  and Kerr effect
at single photon level with the purpose of realizing, in
perspective, a controlled-not gate at single photon level (the
fundamental missing gate for realizing optical quantum computation
\cite{NC}).

\section{Conclusions}

In this proceeding we have presented some recent experiment
realized at the "Carlo Novero" laboratory at IENGF (some other
results is presented in the contribution of M. Chekhova et al. in
this proceedings).

In particular, we have described one of our sources of
polarisation entangled photons, constituted of two $LiIO_3$
crystals whose type I PDC emission (of opposite polarisation,
having rotated the polarisation of the pump laser which pumps both
of them) are superimposed by means of an optical condenser. Since
the distance between the two crystals is smaller of the coherence
length of the pump laser, the emitted state is an entangled one of
the form:
 \be
\vert \Psi \rangle ={\vert H \rangle \vert H \rangle + f \vert V
\rangle \vert V \rangle \over \sqrt{1 + \vert f \vert ^2}}
\label{PsiH2} \ee where the degree of entanglement $f$ can be
tuned by varying the pump intensity and phase between the two
crystals. This source is rather brilliant: we measured a 10 kHz
coincidence rate at 200mW pump power.

Because of these properties this source represents an interesting
device for quantum communication and foundations of quantum
mechanics experiments.

Then, we have reviewed an  experiment realized with it, where
non-maximally entangled states were used for testing Bell
inequalities for the first time. Furthermore, we have described a
double slit experiment that we have realised with a modification
of this set-up, which was addressed to a first comparison between
Standard Quantum Mechanics and de Broglie-Bohm theory, with an
unfavourable result for this last.

We have then reported on a recent experiment about wave-particle
duality that was devoted to overcome limitations of a previous one
\cite{jap}, giving a final answer to the theoretical discussion
started by ref. \cite{P0}

Finally, we have hinted to the realization of a pulsed high
intensity PDC source which will be used in a experiment about
photon-photon interaction in atomic medium.

Altogether we hope to have given an idea of the experimental
activity about Foundations of Quantum Mechanics and Quantum
Information performed in "Carlo Novero" laboratory in our
institute.

\subsection{Acknowledgments}

We would like to acknowledge support by  Italian Minister of
University and Research MURST (contract 2001023718-002 and FIRB
RBAU01L5AZ-002), by INTAS (grant \#01-2122) and by Regione
Piemonte.

\newpage
{\bf Figures Captions} vskip 2cm

Fig. 1 Sketch of the source of polarisation entangled photons.
NLC1 and NLC2 are two $LiIO_3$ crystals cut
   at the phase-matching angle of $51^o$. L1 and L2 are two identical plano-convex lenses with a hole of 4 mm in the
   centre. C is a 5 x 5 x 5 mm quartz plate for birefringence compensation and $\lambda / 2$ is a first order half
   wave-length plate at 351 nm. U.V. identifies the pumping radiation at 351 nm.

   Fig.2 Contour plot of the quantity $CH/N$ (see Eq. \ref{eq:CH}. N is the
total number of detections) in the plane with $f$ (non maximally
entanglement parameter, see the text for the definition) as y-axis
and $\eta$ (total detection efficiency) as x-axis. The polarisers
are supposed to have  $\epsilon _i^{||}=0.99$. The leftmost region
corresponds to the region where no detection loophole free test of
Bell inequalities can be performed. The contour lines are at 0,
0.05, 0.1, 0.15, 0.2.

Fig. 3 Scheme of our experiment where two different pairs produced
in two different point of the crystal are shown. The first pair is
produced exactly on the double slit symmetry axis. The second is
produced 10 $\mu m$ above this, and is at the limit of geometrical
acceptance. The cones represent the spatial dispersion of photons
of the single pair due to finite crystal dimension and pump width.
Figure is not in scale.

Fig.4 Coincidences data in the region of interest compared with
quantum mechanics predictions.
 On the x-axis we report the position of the first detector respect to the median symmetry axis of the double slit.
 The second detector is kept fixed at -0.055 m (the region without data around this point is due to the superposition of the two detectors).
The x errors bars represent the width of the lens before the
detector. A correction for laser power  fluctuations is included.

Fig. 5   Plot of coincidences pattern (in arbitrary units) as a
function of the positions of the first
   photo-detector when the second one is kept fixed at $-1$ cm from the symmetry
   axis.

   Fig.6 The experimental set-up. A vertically polarized Argon laser
beam at 351 nm pumps a lithium iodate crystal (5x5x5 mm) where
type I PDC (i.e. horizontally polarized) is produced. One photon
of the PDC correlated pairs (at 633 nm) is detected, after an
iris, a lens (L) and an interference filter  (IF) by an avalanche
single photon-detector (D3) and used as start of two Time to
Amplitude Converters. To these TACs are  then routed (as stop) the
signals obtained by two single photon detectors (D1 and D2) placed
on the ordinary ($45^o$) and extraordinary ($135^o$) paths
selected by a calcite crystal placed on the conjugated direction
(789 nm) to the former one (both detectors preceded by an iris,
lens  and an interference filter). The outputs of the two TACs are
then routed to an AND circuit giving coincidences ($N_c$).

Fig. 7  Values of the parameter $\alpha$ (see text) for heralded
single photons produced by PDC in function of the single counts of
trigger detector (intensity of the pump laser).

Fig. 8 Experimental set-up for a classical source. The attenuated
classical source light is focalized into the calcite crystal
splitting ordinary and extraordinary rays. The two branches are
than measured by single photon detectors which are routed as stop
to two TACs. The start signal to TACs is given by a pulse
generator. The outputs of TACs feed an AND logical gate giving the
coincidence counts.

Fig. 9 Values of the parameter $\alpha$ in function of the single
counts of one of the detectors for an He-Ne laser beam.

Fig. 10   Values of the parameter $\alpha$ in function of the
single counts of one of the detectors for thermal light (tungsten
lamp emission)

Fig. 11 Sketch of the optical bench with the pulsed high intensity
PDC source. A 355 nm pump beam is injected in a BBO crystal and
produces PDC emission. Two conjugated directions are spatially
selected and focused by lenses in the input slits of
monochromators. The one corresponding to the wavelength 780 nm is
sent into a  Rb gas sample. The two conjugated directions are then
picked up by single photon detection modules and the signals are
routed to a coincidence circuit. The tracing of the chosen
wavelengths in the PDC emission is obtained using a laser diode
beam locked to the Rb absorption line. The locking is achieved
sending the laser beam in a Rb sample, monitoring the absorption
on a photodiode and modifying consequently the length of the
extended cavity of the laser. The polarizing beam splitter in
figure, jointly with a rotatable lambda-quarter waveplate, allows
for directing the beam towards the photodiode or towards the BBO
crystal.

Fig. 12 Picture of the optical bench with the pulsed high
intensity PDC source.
   On the bench one can recognize the Neodimium-Yag laser (foreground right), the BBO crystal,
   the monochromators followed by detectors (background) and the rubidium-line locking system (foreground left).

\vskip 2cm


\begin{thebibliography}{99}

\bibitem{NC}  see for example M.A. Nielsen and I.L. Chuang, { \it Quantum
computation and Information}, Cambridge 2000; D. Bouwmeester et
al., { \it The physics of quantum information}, Springer 2000; N.
Gisin et al., rev. Mod. Phys. 74 (02) 145; M. Keyl, Phys Rep. 369
(02) 431  and ref.s therein.
 \bibitem{tele} C.H. Bennett et al., \PRL { \bf 70}, 1895, 1993. D.
 Boschi et al., \PRL {\bf 80}, 1121, 1998; D. Bouwmeester et al.,
 { \it Nature}, {\bf 390}, 575, 1997;

 \bibitem{dc}C.H. Bennett and S.J. Wiesner, \PRL { \bf 69}, 2881,
 1992. K. Mattle et al., \PRL {\bf 76}, 4656, 1996.
 \bibitem{sw} M. Zukowski et al., \PRL {\bf 71}, 4287, 1993; J.-W. Pan et al., \PRL {\bf 80}, 3891,
 1998.
\bibitem{lontano}  N.Gisen et al., Rev. Mod. Phys. 74 (02) 145 and Ref.s therein.
\bibitem{metr} G. Brida et al., quant-ph/0312211, \PRA in press.


\bibitem{Aul} see for example G. Auletta, {\it Foundations and interpretation of quantum mechanics}
(World Scientific, Singapore 2000) and ref.s therein.

\bibitem{asp}   A. Aspect et al., \PRL { \bf 49} , 1804, 1982.


\bibitem{Mandel}  see L. Mandel, and E. Wolf, Optical Coherence and Quantum Optics,
Cambridge University Press, 1995 and references therein.
\bibitem{franson}    J. P. Franson, \PRL { \bf 62}, 2205, 1989.
\bibitem{ou}    Z.J. Ou and L. Mandel, \PRL { \bf 61}, 50, 1988;
Y.H.Shih et al., \PRA { \bf 47}, 1288, 1993.
\bibitem{type1}   J. G. Rarity, and P. R. Tapster, \PRL { \bf 64},
2495, 1990; J. Brendel et al., {\it Eur.Phys.Lett.} { \bf 20},
275, 1992; P. G. Kwiat el al., \PRA { \bf 41}, 2910, 1990; W.
Tittel et al., \PRL { \bf 81}, 3563, 1998.

\bibitem{type2}    T.E. Kiess et al., \PRL { \bf 71}, 3893, 1993;
P.G. Kwiat et al., \PRL { \bf 75}, 4337, 1995.

\bibitem{whi} A. G. White et al., \PRL { \bf 83}, 3103 1999;
\bibitem{shih}Y. Kim et al., \PRA { \bf 63}, 060301(R), 2001; \PRA { \bf 63}, 062301,
2001.

\bibitem{nos} G. Brida, M. Genovese,
C. Novero and E. Predazzi, \PLA { \bf 268}, 12, 2000; Proc. of
QCM\&C 3 (Capri 2000), ed. P. Tombesi and O. Hirota, Kluwer,
p.399.

\bibitem{kw} P.G. Kwiat et al., \PRA { \bf 60}, R773, 1999.


\bibitem{bell} J.S. Bell, { \it Physics} { \bf 1 }, 195, 1965 .
\bibitem{santos}  E. Santos, \PLA { \bf 212}, 10, 1996; L. De Caro and A.
Garuccio \PRA { \bf 54}, 174, 1996 and references therein.

\bibitem{BK} P.H. Eberhard, Nucl. Phys. B { \bf 398}, 155, 1993; A. Di Domenico,  Nucl. Phys. B { \bf 450},  293, 1995;
B. Ancochea, \PRD { \bf 60}, 094008, 1999; A. Bramon and M.
Nowakowski, \PRL { \bf 83}, 1, 1999; N. Gisen and A. Go, { \it
Am.J.Phys.} {\bf 69}, 264, 2001; B.C. Hiesmayr, { \it
Found.Phys.Lett.} { \bf 14}, 231, 2001; P. Privitera and F.
Selleri, \PLB { \bf 296}, 261, 1992; F. Selleri, \PRA { \bf 56},
3493, 1997; A. Pompili and F. Selleri, {\it Eur. Phys. Journ. C} {
\bf 14}, 469, 2000; A.Bramon and G.Garbarino, \PRL { \bf 88},
040403, 2002; A.Bramon and G.Garbarino, \PRL { \bf 89}, 160401,
2002.
\bibitem{Win} M. A. Rowe et al., Nature { \bf 409}, 791, 2001.
\bibitem{BGN} G. Brida, M. Genovese and C. Novero,{ \it Eur. Jour. of Phys. D.} { \bf 8},
273, 2000.
\bibitem{nosK} M.Genovese et al., \PLB { \bf 513}, 401, 2001; Foud. Of Phys. { \bf 32}, 589, 2002.
M. Genovese, Phys. Rev A 69, (2004) 022103.

\bibitem{eb}  P. H. Eberhard, \PRA { \bf 47}, R747, 1993.
\bibitem{Mig} A. Migdall, { \it Phys. Today}, 41, 1999 and references therein.
\bibitem{Santos2} A. Casado et al., quant-ph 0202097.

\bibitem{Santosv} A. Casado et al., J. Opt. Soc. Of Am. B { \bf 14}, 494, 1997; \PRA { \bf 55}, 3879, 1997;
\PRA { \bf 56}, 2477, 1997 ;  J. Opt. Soc. Of Am. B { \bf 15},
1572, 1998; { \it Eur. Phys. Journ. D}, { \bf 11 }, 465, 2000; D {
\bf 13},  109, 2001; E. Santos personal communication
\bibitem{JMO} G. Brida, M. Genovese and C. Novero, { \it Journ. Mod. Opt.} { \bf 47}, 2099, 2000 and ref.s therein.
\bibitem{nosW} G. Brida et al., \PLA {\bf 299}, 121, 2002;
\bibitem{SnosS} G. Brida
et al., {\it Journ. Mod. Opt.} 11 (2003) 1757.
\bibitem{epjd} M. Genovese and C. Novero, { \it Eur.  Journ. of Phys. D.} { \bf 21}, 109, 2002.
\bibitem{s3} E. Santos, quant-ph 0401003.
\bibitem{qutrit} H. Bechmann-Pasquinucci and W. Tittel, \PRA { \bf
61}, 062308, 2000; H. Bechmann-Pasquinucci and A. Peres, \PRL {
\bf 85 }, 3313, 2000; M.Bourennane et al., \PRA { \bf 63}, 062303,
2001; N.J. Cerf et al., \PRL { \bf 88}, 127902, 2002; D. Bruss and
C. Macchiavello, \PRL { \bf 88}, 127901, 2002; D.B. Horoshko and
S.Y. Kilin, quant-ph 0203095; M. Genovese "On limit quantum
efficiency for violation of Clauser-Horne Inequality for qutrits",
in press.

\bibitem{Kerr} D. Vitali et al., \PRL { \bf 85}, 445, 2000; M. Genovese, \PRA {\bf 63}, 044303, 2001;
C. Ottaviani et al., \PRL { \bf 90}, 197902, 2003; M. Genovese and
C. Novero, Phys. Lett. A 271 (2000) 48..

\bibitem {4} P. Ghose, in {\it Foundations of quantum theory and quantum optics 1999/2000}(ed. Roy, S.M.) (Indian Academy of Sciences) 211.

\bibitem {5} P. Ghose, A.S. Majumdar, S. Guha,  and J. Sau , \PLA {\bf 290}, 205, 2001; quant-ph 0103126.
\bibitem {6} M. Golshani, and O. Akhavan , J. Phys. {\bf A34}, 5259, 2001.
\bibitem {7} see for example P. Ghose , {\it Testing quantum mechanics on a new ground} (Cabridge Univ. Press, 1999) and ref.s therein.
\bibitem{disc} L. Marchildon, quant-ph 0007068; quant-ph 0101132; P. Ghose, quant-ph 000807; quant-ph 0103136; private communication ; W. Struyve and W. De Baere, quant-ph 0108038; P. Holland, private communication. ; P. Ghose quant-ph 0208192.
\bibitem{nosdBB} G. Brida et al., { \it J. Phys. B: At. Mol. Opt. Phys.} { \bf 35}, 4751,
2002; Phys. Rev. A 68 (2003)033803.
\bibitem{news} J. Mullins, Newscientist July 20th 2002, 21.
\bibitem{joe} A. Joobeur et al., \PRA { \bf 54}, 3349, 1994.
\bibitem{LA}  H. Bechmann-Pasquinucci and W. Tittel, {Phys. Rev.} A { \bf
61}, 062308, 2000; H. Bechmann-Pasquinucci and A. Peres, \PRL {
\bf 85 }, 3313, 2000; M.Bourennane et al., \PRA { \bf 63}, 062303,
2001; N.J. Cerf et al., \PRL { \bf 88}, 127902, 2002; D. Bruss and
C. Macchiavello, \PRL { \bf 88}, 127901, 2002.
\bibitem{Gisen}  J. Brendel et al., \PRL { \bf 82}, 2594, 1999; W.
Tittel et al., Phys. Rev. Lett. { \bf 84}, 4737, 2000.


\bibitem{Bohr}N. Bohr, Atomic Theory and description of Nature,
Cambridge Univ. press 1934.

\bibitem{P1} P. Ghose, D. Home and G.S. Agarwal, \PLA 168 (1992)
95.


\bibitem{PL} P. Ghose, Testing quantum mechanics on a new ground,
Cambridge Univ. Press (1999) and ref.s therein.

\bibitem{int} see for example: P. Grangier, G. Roger and A. Aspect, Eur. Phys. Lett. 1
(1986) 173; A. Aspect and P. Grangier, Hyp. Int. 37 (1987) 3.

\bibitem{P0} P. Ghose, D. Home and G.S. Agarwal, \PLA 153 (1991)
403.

\bibitem{jap} Y. Mizobuchi and Y. Ohtak\'e, \PLA 168 (1992) 1;

\bibitem{Un} C.S. Unnikrishnan and S.A. Murthy, \PLA 221 (1996) 1.

\bibitem{noswp} G.Brida, M.Genovese, M.Gramegna and E. Predazzi.
\PLA 328 (04) 313.
\bibitem{santosM} T.W. Marshal and E. Santos, Found. of Phys. 18
(1988) 185 and in "Problems in quantum Physics", eds. L. Kostro et
al., (World Scientific, Singapore, 1988).

\bibitem{atom} S. Lloyd et al., quant-ph 0003147.

\bibitem{or} R. Jozsa et al., \PRL {\bf 85}, 2010, 2000; M.Genovese and C. Novero, quant-ph 0009119.

\end{thebibliography}
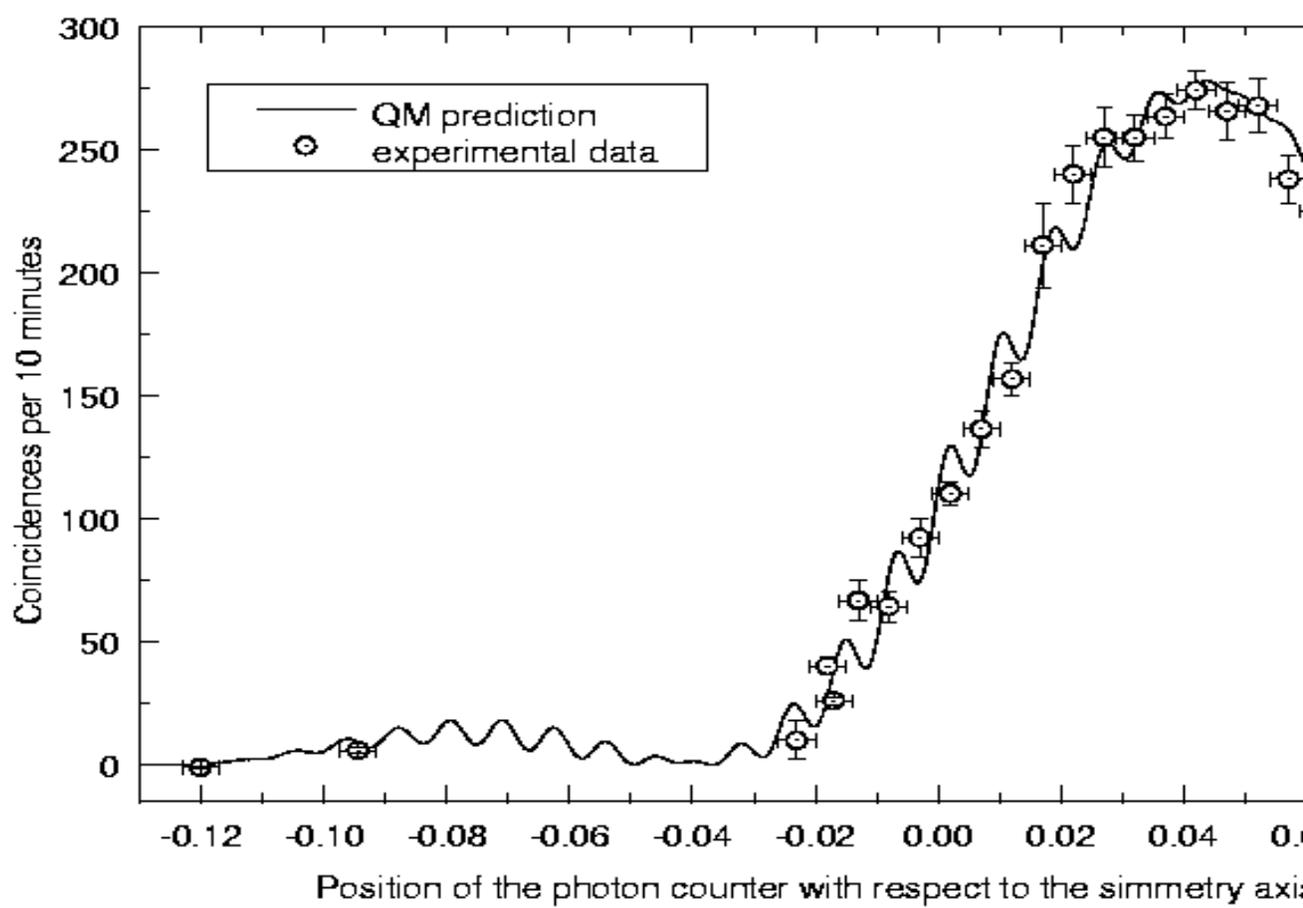
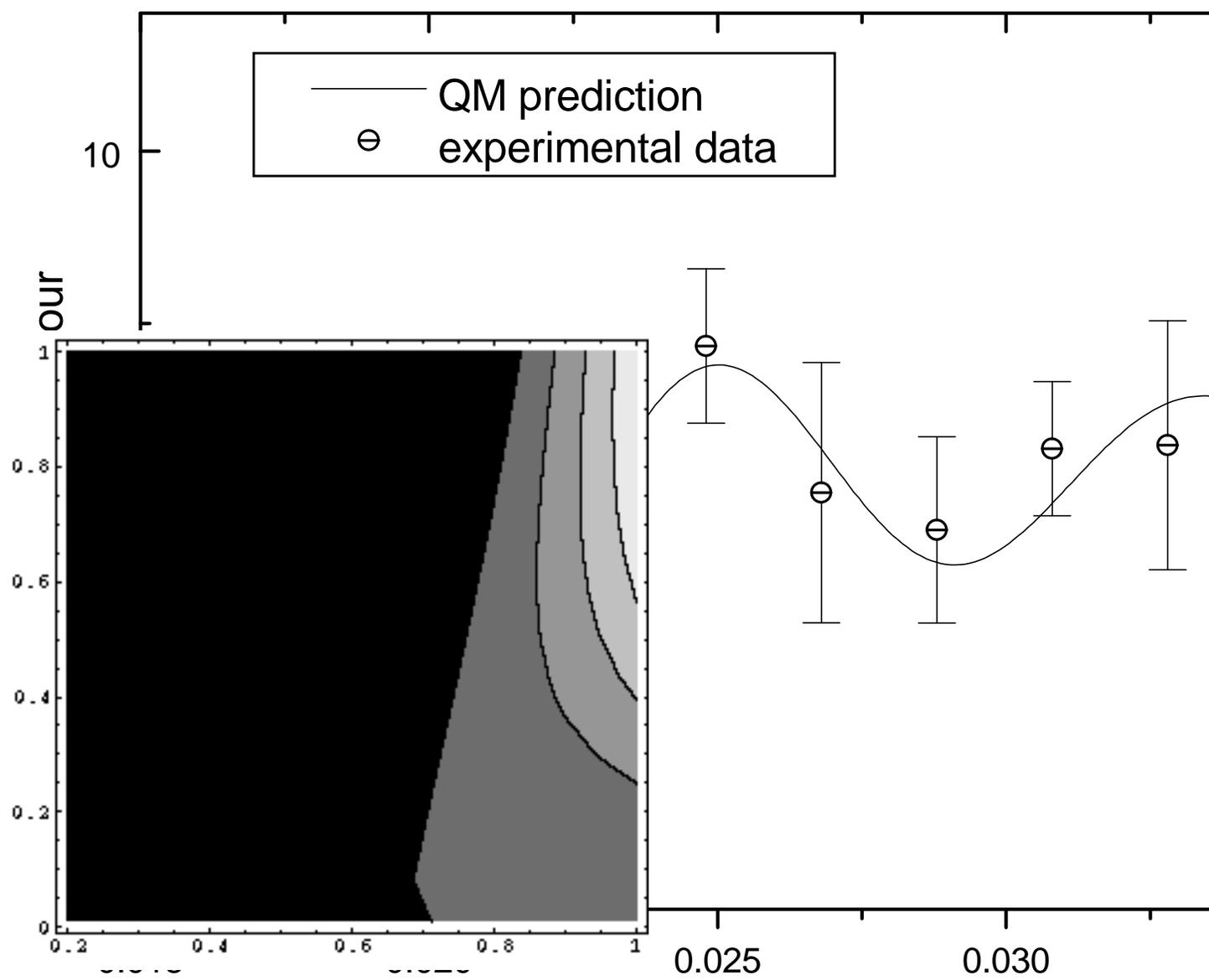
\end{document}